\documentclass[twocolumn]{aastex62}

\usepackage{graphicx}   
\usepackage{amsmath}    
\usepackage{amssymb}    
\usepackage{comment}
\usepackage{xspace}
\usepackage{lastpage}

\usepackage{eso-pic}

\usepackage{comment}

\def\reff@jnl#1{{\rm#1\/}}
\def\aj{\reff@jnl{AJ}}         
\def\araa{\reff@jnl{ARA\&A}}      
\def\apj{\reff@jnl{ApJ}}        
\def\apjl{\reff@jnl{ApJ}}        
\def\apjs{\reff@jnl{ApJS}}       
\def\aap{\reff@jnl{A\&A}}        
\def\aapr{\reff@jnl{A\&A~Rev.}}     
\def\aaps{\reff@jnl{A\&AS}}       
\def\mnras{\reff@jnl{MNRAS}}      
\def\physrep{\reff@jnl{Physics Reports}}
\def\prd{\reff@jnl{Phys.Rev.D}}     
\def\prl{\reff@jnl{Phys.Rev.Lett}}   
\def\pasp{\reff@jnl{PASP}}       
\def\pasj{\reff@jnl{PASJ}}       
\def\nat{\reff@jnl{Nature}}       
\def\jcap{\reff@jnl{JCAP}}   
\def\memsai{\reff@jnl{MemSAI}} 
\def\na{\reff@jnl{New Astronomy}}       
\def\procspie{\reff@jnl{SPIE}}       
\def\pasa{\reff@jnl{PASA}}

\begin{document}

\title{Planet X in CMB and Optical Galaxy Surveys}

\author{Eric J. Baxter}
\affiliation{Department of Physics and Astronomy, University of Pennsylvania, Philadelphia, PA 19104, USA}\email{ebax@sas.upenn.edu}

\author{Bhuvnesh Jain}
\affiliation{Department of Physics and Astronomy, University of Pennsylvania, Philadelphia, PA 19104, USA}

\author{Cullen H. Blake}
\affiliation{Department of Physics and Astronomy, University of Pennsylvania, Philadelphia, PA 19104, USA}

\author{Gary Bernstein}
\affiliation{Department of Physics and Astronomy, University of Pennsylvania, Philadelphia, PA 19104, USA}

\author{Mark Devlin}
\affiliation{Department of Physics and Astronomy, University of Pennsylvania, Philadelphia, PA 19104, USA}

\author{Gil Holder}
\affiliation{Department of Physics and Astronomy, University of Illinois at Urbana-Champaign, IL 61801, USA}

\defcitealias{Cowan:2016}{CHK}

\begin{abstract}
 We consider the possibility of detecting and tracking the hypothesized Planet 9 or other unknown planetary-mass distant solar system members, generically called Planet X, with a combination of CMB and optical imaging surveys. Planets are detectable via thermal emission in CMB surveys and via reflected sunlight in optical surveys. Since the flux from reflected light falls off faster with distance, the signal-to-noise of planetary observations with optical  surveys falls off faster than for CMB surveys.  A promising approach to detecting new solar system planets with future surveys such as the Simons Observatory, CMB-S4 and LSST, is for a detection in CMB data followed by tracking in the synoptic imaging survey. Even if the parallax were not detected in CMB data,  point sources consistent with thermal spectra could be followed up by LSST. In addition to expanding the Planet X discovery space, the joint datasets would improve constraints on key orbital and thermal properties of outer solar system bodies. This approach would work for a Neptune-like planet up to distances of a few thousand AU, and for an Earth-like planet up to several hundred AU. We discuss the prospects for the next decade as well as nearer-term surveys. 
\vspace{1cm}
\end{abstract}

\section{Introduction}

The existence of one or more undetected Earth or Neptune-sized planets in our solar system with large perihelion distance has been suggested in recent years by several authors. \citet{Trujillo:2014}, \citet{Batygin:2016}, and \citet{Malhotra:2016} have made the case for a Planet 9 with mass of order 10 Earth masses and distance of order 700 AU.  \citet{Brown:2016} constrain the mass of such a planet to 5 to 20 Earth masses, and its semi-major axis to between 380 and 980 AU.  \citet{Volk:2017} make the case for an additional planetary object with mass 0.1 to 2.4 Earth masses and a distance of 60 to 100 AU.   In general, the presence of additional planets beyond 1000 AU is poorly constrained.  We will refer to any undetected planet in the solar system generically as Planet X. 

Searches in the optical for detecting Planet 9 via reflected sunlight are underway with wide-field instruments such as the Subaru Hyper Suprime-Cam, Pan-STARRS and the Dark Energy Survey with the Blanco telescope  \citep[e.g.][S. More, private communication]{Sheppard:2018,Weryk:2016,Pedro:2019}.  \citet{Cowan:2016} (henceforth \citetalias{Cowan:2016}) note that for reasonable assumptions, the expected temperature of Planet 9 is in the range 28-53 Kelvin, so the peak thermal emission is at submillimeter wavelengths and natural datasets to pursue this signal are surveys designed to map the Cosmic Microwave Background (CMB) at millimeter wavelengths.  They also note that at an approximate distance of 700~AU, the parallax motion of Planet 9 would be a few arcseconds per day (or about 10 arcminutes per six months), which would dominate over its proper motion. So a CMB survey with roughly arcminute resolution that repeats coverage of the same part of sky every few months could potentially detect Planet 9 as a moving source. The sensitivity and angular resolution requirements are discussed in detail by \citetalias{Cowan:2016}, who note that such a planet would be at the edge of detectability by {\it Planck}. 

We compare the signal-to-noise of Planet X measurements in optical vs CMB surveys and consider the  possibility that such a planet could be detected in CMB data (via its thermal emission) and tracked in optical surveys (via its reflected light), or other useful combinations of the two types of surveys. The surveys have differences in several key characteristics: sensitivity, resolution, cadence, and confusion with other sources. 
 
Beyond ongoing galaxy surveys, for the 2020's we  focus on the Large Synoptic Survey Telescope (LSST), while noting the potential strengths of the space-based surveys with Euclid, SPHEREx and WFIRST. For the CMB, the South Pole Telescope (SPT) and Atacama Cosmology Telescope (ACT) projects have ongoing surveys. In the 2020's, the planned Simons Observatory (SO) and CMB-S4 telescopes will carry out wide area surveys with greater sensitivity. As we discuss below, the key issue for Planet X detection, beyond the known sensitivity and resolution of the surveys, is their cadence.  

The paper is organized as follows: in \S\ref{sec:model} we describe the expected properties of Planet X and related bodies; in \S\ref{sec:surveys} we describe ongoing and planned surveys; in \S\ref{sec:prospects} we describe strategies for detecting and tracking Planet X.

\section{Modeling}
\label{sec:model}

The temperature of planets is set by equilibrium between absorption of solar and interstellar light, internal heat from sources such as radioactive decays and primordial heat, and thermal emission from the objects.  The material properties and sizes of planets determine how efficiently they absorb and emit radiation, and therefore their equilibrium temperatures.  The brightness in reflected sunlight depends on the albedo of the planets.

Assuming a temperature of $T_{\rm bg} = 3.5\,{\rm K}$ for the background temperature of interstellar light and the CMB, we have
\begin{eqnarray}
T(r) = \left( T_{\rm bg}^4 + \frac{(1-A) L_{\odot}}{16\pi \sigma r^2} + \frac{P_{\rm int}}{4\pi\sigma R^2} \right)^{1/4},
\end{eqnarray}
where $A$ is the Bond albedo, $r$ is the star-planet distance, $R$ is the planetary radius, and $\sigma$ is the Stefan-Boltzmann constant.  The Bond albedo of both Earth and Neptune is approximately 0.3, and we will adopt that value here.  We have assumed that the emissivity of the planet is $\epsilon = 1$, independent of wavelength. 

For Earth, the internal power is well constrained to be $P_{\rm int} = 47\times 10^{12}\,{\rm W}$.  For Planet 9, we follow \citetalias{Cowan:2016}, and assume an internal power ranging from $3\times 10^{14}\,{\rm W}$ to $3\times 10^{15}\,{\rm W}$, since this is roughly the range spanned by Uranus and Neptune. Given the estimated mass of Planet 9 (i.e. greater than $10M_{\Earth}$, \citealt{Batygin:2016}), statistical analyses of large populations of exoplanets indicate that we should expect Planet 9 to have a density, and possibly composition, similar to that of the known gas giant planets (\citealt{Weiss:2013}). Our baseline assumption is that its properties are similar to Neptune. 

\begin{figure}[th]
\centering
\includegraphics[width=9cm]{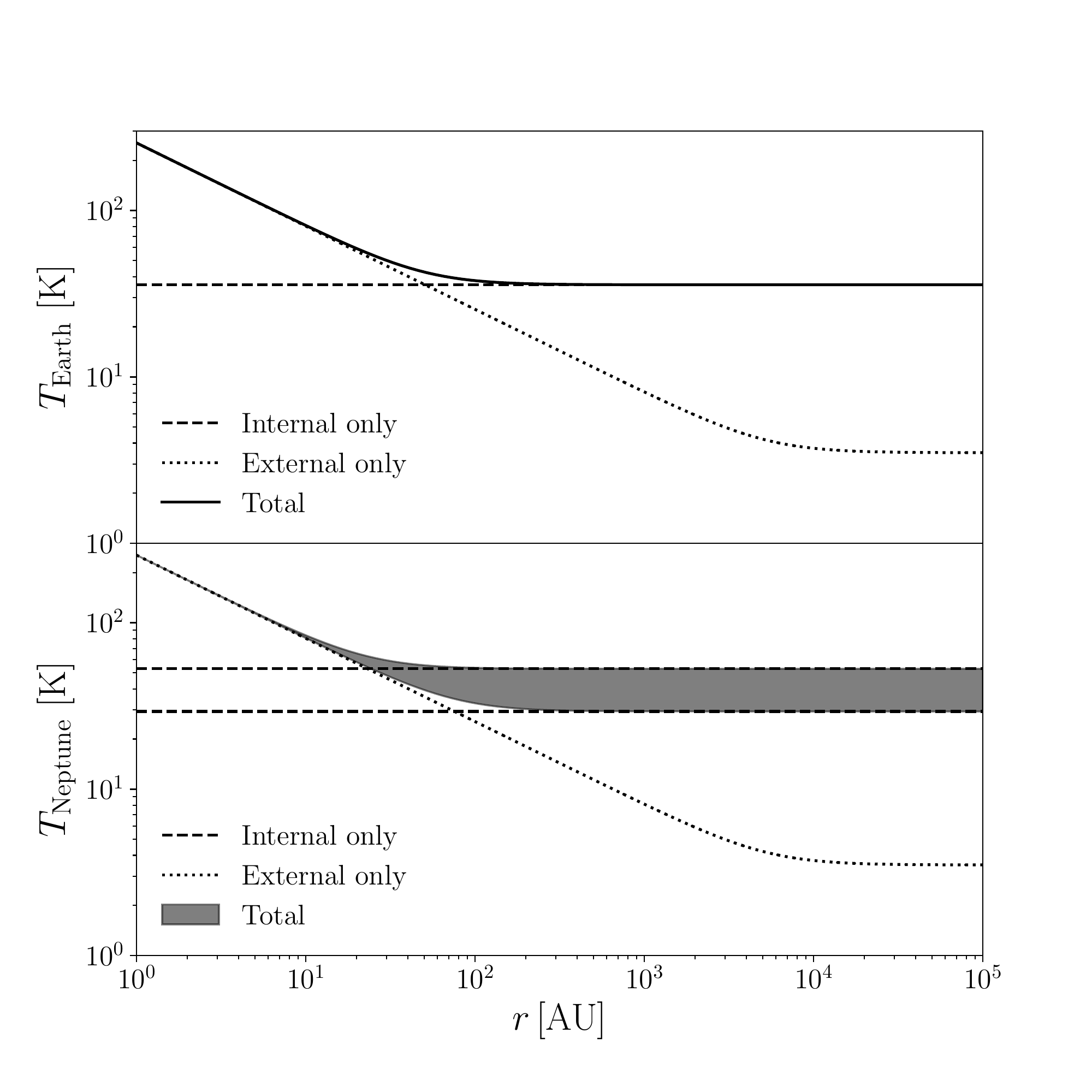}
\caption{
Temperature of Earth-like (top) and Neptune-like (bottom) planets as a function of distance from the sun.  Solid regions represent total temperature, while dotted and dashed lines represent temperatures given only external and internal energy sources, respectively. The width of the total temperature band for the Neptune-like planet represents the uncertainty on internal heat input for Neptune.  Beyond roughly 200 AU from the sun, a planet's temperature is dominated by internal heat sources.  
}
\label{fig:planet_temperature}
\end{figure}

As shown in Figure~\ref{fig:planet_temperature}, beyond about $r \gtrsim 200\,{\rm AU}$, the temperature of an Earth or Neptune-like planet is dominated by internal heat sources, and is therefore roughly constant with distance from the sun.  We will assume this limit in the analysis below.  Consequently, for an Earth placed in the outer solar system, we adopt a temperature of 35 K.  For Planet 9, plausible temperatures are in the range of 28 to 53 K; we will adopt 50 K as our fiducial value.   Note that the contribution to the planet's temperature from interstellar light and the cosmic background radiation are negligible below about $10^4$ AU in comparison to the sun.

Detecting reflected sunlight from objects in the outer solar system is challenging since the strength of the signal drops off as $r^{-4}$. The apparent brightness of a planet depends on the radius of the planet, as well as the optical properties of its surface. Neglecting the distance between the sun and Earth, the observed flux from the planet at wavelength $\lambda$, $F_{\lambda}$, is
\begin{equation}
    F_{\lambda} = a_{\lambda} \frac{L_{\lambda}R^{2}}{4\pi r^{4}},
\end{equation}
where $L_{\lambda}$ and $a_{\lambda}$ are the solar luminosity and the geometric albedo of the planet, respectively, at the wavelength of interest. In our solar system, 
the geometric albedo across the optical and infrared range widely, over at least a factor of five. Given that we expect Planet X or another large outer solar system body to have a gaseous, low-density composition, a geometric albedo similar to that of Neptune or Uranus is a reasonable assumption ($a_{\lambda}\approx0.4)$.  While some exoplanets have been found to have exceptionally low albedos (e.g. \citealt{Rowe:2008}), these are hot Jupiter planets where the atmospheric chemistry would be very different from that of Planet X.

\begin{figure}[th]
\centering
\includegraphics[width=9cm]{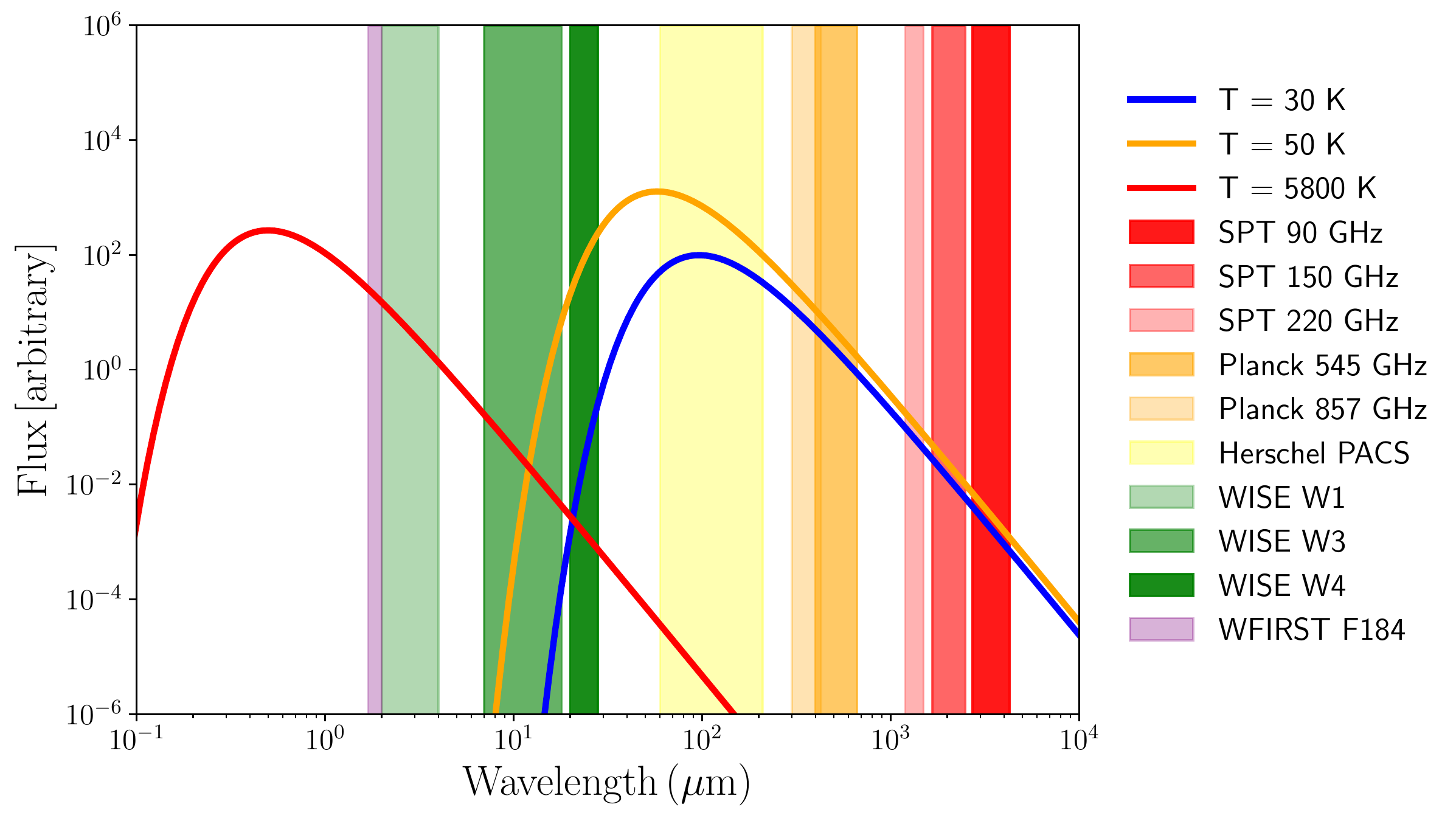}
\caption{
Spectra of thermal emission ($T= 30\,{\rm K}$ and $T= 50\,{\rm K}$) and reflected solar emission ($T= 5800\,{\rm K}$) for Planet X, compared to observing bands from several experiments.
}
\label{fig:bandpasses}
\end{figure}

\begin{table*}
    \centering
    \begin{tabular}{|c|c|c|c|c|c|}
    \hline
         Survey name & Type & Years active & Sky area [deg.$^2$]& Sensitivity  & Resolution (FWHM) \\ \hline
         DES    & Optical & 2013-2019 & 5000   & 24 in $r$-band & 1\arcsec \\ \hline
         HSC    & Optical & 2014-2020 & 1400   & 25 in $r$-band & 0.8\arcsec \\ \hline
         LSST   & Optical & 2023-2033 & 18,000 & 24.5 in $r$-band & 0.8\arcsec \\ \hline
         SPT-SZ & CMB     & 2008-2011 & 2500   & 3.4 mJy  & 1\arcmin \\ \hline
         SO     & CMB     & 2020-2023 & 16,000  & 1.7 mJy & 1.4\arcmin \\ \hline
         CMB-S4 & CMB     & 2025-2030 & 20,000 & 0.4 mJy & 1.4\arcmin \\
         \hline
         
    \end{tabular}
    \caption{Summary of current and upcoming CMB and optical imaging surveys considered in this analysis. Note that the sensitivity for optical surveys is for a single exposures. For LSST the sensitivity of the full survey co-adds ($r=27.5$) is significantly better; see also discussion of the sky coverage in \citet{Trilling:2018}.  For the CMB surveys, we have reported the sensitivities and beam sizes at 150 GHz. 
    }
    \label{tab:survey_properties}
\end{table*}

\bigskip
\bigskip
\bigskip

\section{Planet X detection capabilities of current and planned Surveys}
\label{sec:surveys}

We summarize the properties of various current and future optical and CMB surveys in Table~\ref{tab:survey_properties}.  The {\it Planck} satellite mapped the CMB in nine  frequency channels and has released its completed survey data. Its highest  frequency channel is centered at 857 GHz. The wavelength corresponding to 857 GHz, 0.35~mm, is close to the peak of thermal emission  at a temperature of $\sim 10K$. Ground-based CMB surveys have a smaller frequency coverage and a maximum frequency typically well below that of {\it Planck}, so these surveys typically cover the Rayleigh-Jeans part of the thermal spectrum for objects with temperature above 20K. 
Since CMB surveys are usually optimized for frequencies near 150 GHz, and ongoing surveys have much better sensitivity than {\it Planck}, we will concentrate on those frequencies here (though the signal from a $\sim 30$~K blackbody peaks at frequencies of 1800 GHz, the noise also increases with frequency).  Figure~\ref{fig:bandpasses} summarizes the frequency coverage of several current and future surveys relative to both the thermal emission and reflected light (i.e. $T = 5800\,{\rm K}$) from a Planet X.

Among ongoing CMB experiments,  SPT  \citep{Carlstrom:2011} and ACT \citep{Swetz2011} offer higher resolution and deeper maps than {\it Planck} over thousands of square degrees of the sky.  Both experiments have completed multiple wide area CMB surveys \citep{Story:2013, Sievers:2013}.  Upgraded versions of the original SPT and ACT receivers --- SPT-3G \citep{Benson:2014} and Advanced ACTPol \citep{Henderson:2016} --- are currently performing deeper surveys.  The planned Simons Observatory \citep{Galitzki:2018} will be on the sky in the early 2020's (a similar timeframe as LSST) and its survey will improve on ongoing surveys in sensitivity and sky coverage. It will serve as a precursor to the planned CMB-S4 experiment, scheduled for the mid-2020's \citep{Abazajian:2016}. Several other telescopes are being planned \citep[see e.g.][]{Stevens:2018}; we do not discuss these further but note that CCAT-prime may be promising for Planet X  studies depending on its planned survey \citep{Stacey:2018}. The northern sky  is sparsely covered by the CMB experiments, which are mostly based in Chile or at the South Pole, leaving a significant gap in the reach of CMB surveys. However, LSST is also based in Chile, so the survey footprints have good overlap. 

\begin{figure}[th]
\centering
\includegraphics[width=8.5cm]{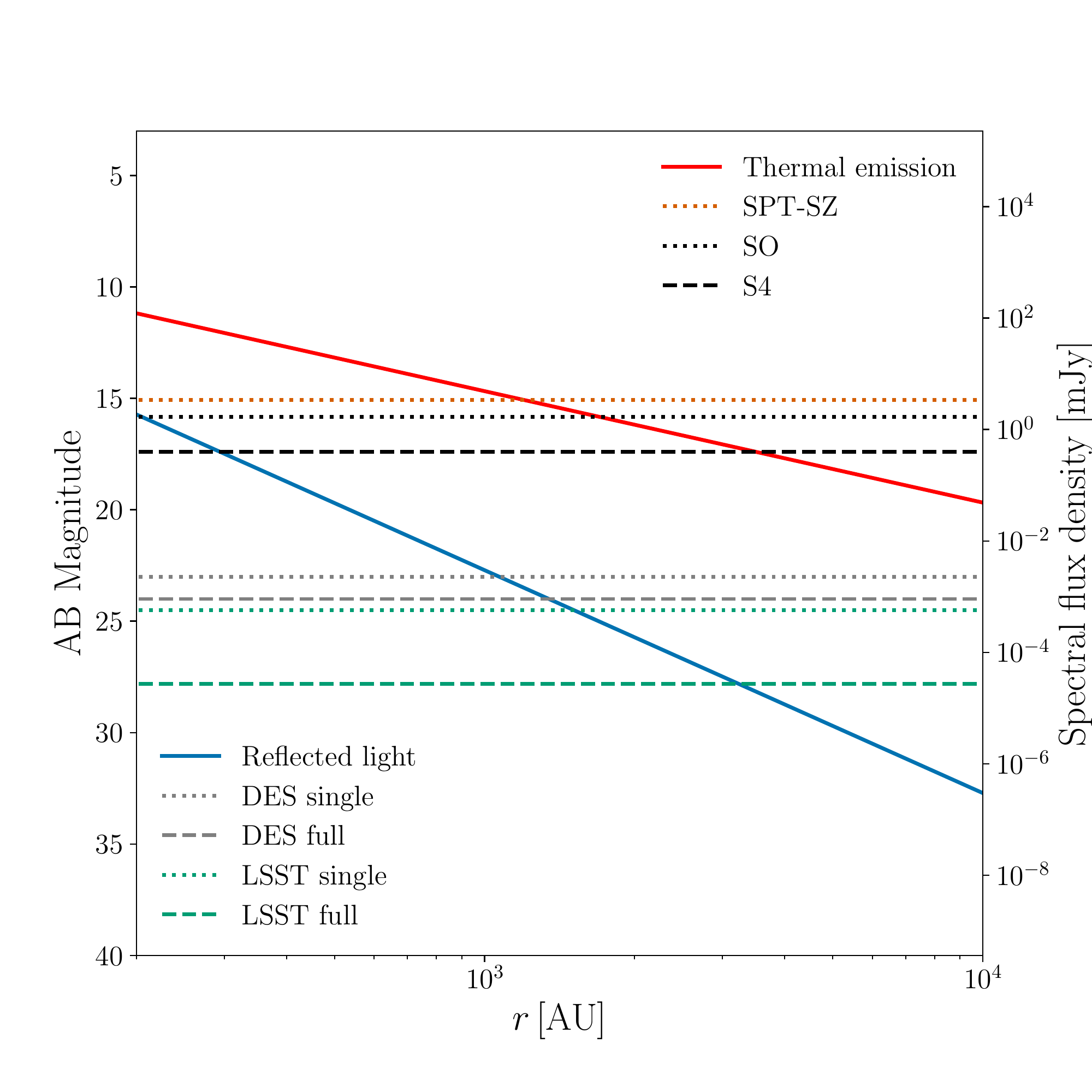}
\caption{
The signal in thermal emission (red) and reflected light (blue) from a hypothetical  Neptune-like planet as a function of distance from the sun.  Also plotted are $5\sigma$ noise point source detection limits for several current and upcoming galaxy and CMB surveys.
}
\label{fig:cmb-galaxy}
\end{figure}

Ongoing optical surveys include DES, which upon its imminent completion will have imaged each part of its 5,000 deg$^2$ footprint 10 times in each of five bands, with these exposures spread over the six years of survey operations \citep{Flaugher:2005}. The LSST survey will cover the Southern sky \citep{LSST:2012} from Cerro Pachon in Chile. It will take two 15 second exposures per night (called a single visit), and return to each field every few days. Over ten years, each field will have about 100 exposures in each of six filters. \citet{Trilling:2018} consider a 45 day period per observing season for a given field, and require five visits for a useful orbital arc to be  measured. We do not consider in any detail the surveys planned for the space-based telescopes Euclid and WFIRST, but note that they are complementary to LSST. Their infrared coverage expands the wavelength range available, and the superior resolution from space can help with source identification. WFIRST also has excellent astrometric capabilities, which offers another route to planet detection via its lensing of background stars as explored in a recent study by \citet{vanTilburg:2018}, who also discuss the prospects with {\it Gaia} and the proposed {\it Theia} mission.

\begin{figure*}[th]
\centering
\includegraphics[width=8.5cm]{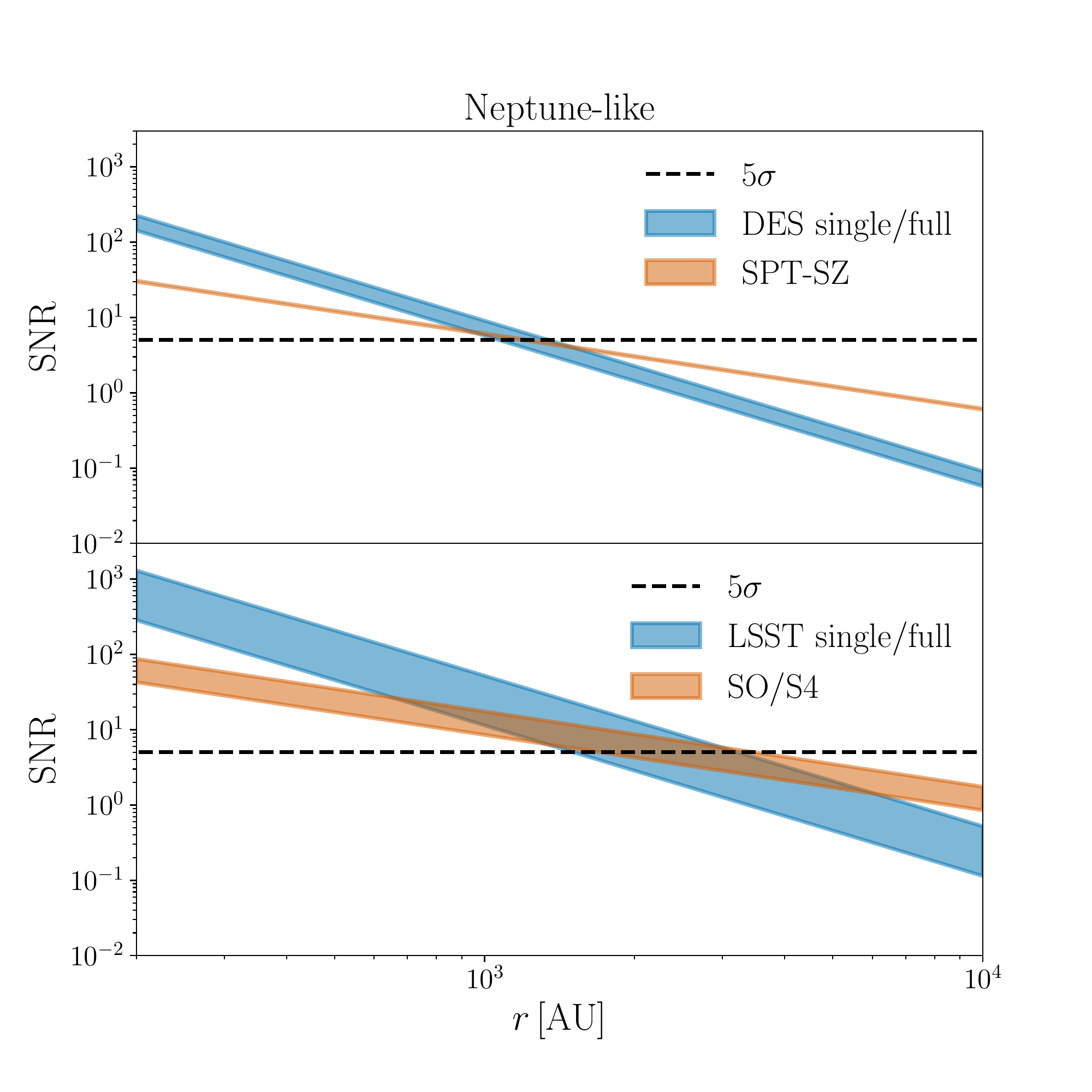}
\includegraphics[width=8.5cm]{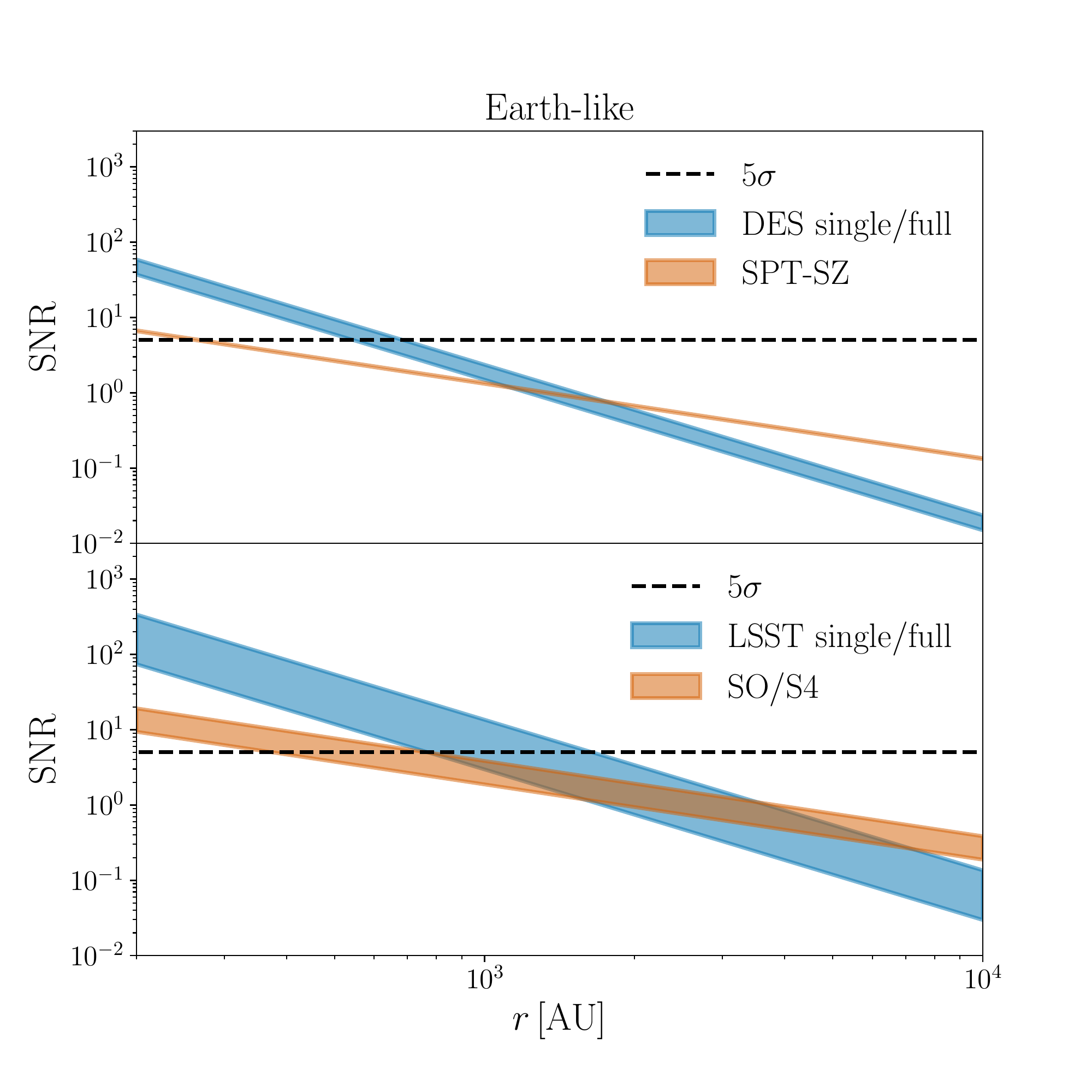}
\caption{
The signal-to-noise ratio of planetary detections for galaxy (blue) and CMB surveys (orange) is plotted as a function of the planet's distance from the Sun. The left panels are for a Neptune-like planet and the right panels for an Earth-like planet. The upper panels show two overlapping completed surveys (DES and the SPT-SZ survey), while the lower panels show surveys that will operate in the 2020's: Simons Observatory (SO), CMB-S4 (S4) and LSST.  The bands for the optical surveys represent the two extremes of using single epoch data or full-survey coadds for detection. The band for SO/S4 indicates the signal-to-noise range for these two surveys.
}
\label{fig:signal-noise}
\end{figure*}

Figure \ref{fig:cmb-galaxy} shows the signal from Planet X  with optical and CMB surveys as a function of distance from the sun. The $5\sigma$ noise levels for several surveys are also shown. A comparison of the signal-to-noise from optical and CMB surveys is shown in Figure~\ref{fig:signal-noise} and discussed in the next section. 

Whether a planet can be distinguished from other point sources depends on the ability to measure its motion with either CMB or optical data.  The parallax motion of a planet at distance $d$ from the sun is roughly
\begin{equation}
\theta = \frac{10\,{\rm arcmin}}{6\,{\rm months}} \left( \frac{700 {\rm AU}}{d} \right).
\end{equation}
For a CMB survey with beam size of roughly 1.5 arcminutes, the parallax motion of a planet much beyond $4700$ AU will be less than a beam size.  For a beam with standard deviation $\sigma_{\rm beam}$, the centroid of the beam can typically be localized to roughly $\sigma_{\rm beam}/{\rm SNR}$, where SNR is the signal-to-noise of the detection.  Assuming a $5\sigma$ detection, parallaxes as small as 0.2 arcmin could be resolved by current and future CMB surveys.  However, several factors may limit the ability of CMB surveys to reach this level of precision.  First, the survey scanning strategy of CMB surveys may mean that the time period over which a given patch of the sky is observed is significantly less than 6 months.  Second, if the number of detected sources is large, reducing the false positive rate of parallax detection to an acceptable value will require centroid separation at several times the centroiding error.  Finally, if difference maps formed from first-half and second-half season maps are used to identify moving sources, the effective survey depth will be reduced by $\sqrt{2}$.  For these reasons, we will assume below that the parallax motion must be roughly $\sigma_{\rm beam}$ in order to be robustly detected in the CMB survey.

If a moving point source is directly detected in CMB data (e.g. via difference imaging), \citetalias{Cowan:2016} argue that the amplitude of its parallax will uniquely identify the source as a promising Planet X candidate.  Once identified, the candidate can be followed up with targeted observations.  If the CMB data has localized a source to 1 arcminute, then a deep observation over this field with an 8m-class telescope could reach LSST full-survey depth in a single night.  The motion will be slow enough to easily stack exposures for a full night.  A repeat observation some nights later will detect the slow-moving source in the field, using image subtraction if necessary. Over a longer time period, the much better resolution of optical data can be used to model the orbital motion in addition to the parallax.  In this case, the main role of a wide-field optical survey may be in providing past data to readily determine the orbit, if the survey has been in progress prior to the detection. 

Similarly, searches for Planet X may be relatively straightforward for optical surveys if the target is bright enough to be detected in a single visit.  Planet X's motion would be small enough to be negligible within a single visit, so it would resemble a point source in single-visit catalogs.  The parallax motion at opposition exceeds 1\arcsec\ day$^{-1}$ at $d<3500$~AU, and thus Planet X should be readily recognized as a slowly moving source in the collection of single-epoch catalogs.  Such searches are underway with Subaru, Pan-STARRS and DES data, and we can expect that they will be executed for LSST and other optical wide-area surveys.  \citet{Trilling:2018} find that a Neptune-sized planet in the LSST footprint would have parallax measurements out to 1000 AU, based on source detection in single visits. 

With an optical identification in hand, mm-wave follow-up can be done with pointed observations from large-aperture telescopes such as ALMA, and the wide-field capabilities of CMB surveys do not necessarily yield an advantage.

\section{Prospects with the combination of CMB and optical surveys}
\label{sec:prospects}

In this section we discuss how the use of both optical and CMB  surveys enlarges the discovery space for Planet X and enables better measurement of its orbital motion and physical properties. 

Figure~\ref{fig:signal-noise} shows the signal-to-noise of these measurements: it highlights the rapid decline with distance ($r^{-4}$) of detection prospects with optical surveys. In terms of the absolute signal-to-noise, the right-hand panels suggest that an Earth-like Planet X that is detectable to CMB surveys would also be bright and close enough to be discovered in single-epoch catalogs for contemporary optical surveys. For a larger Planet X, however, the left-hand panels show that a Neptune-sized planet at distance below 1000 AU is accessible to both types of surveys, but the CMB surveys will significantly extend the discovery distance of Planet X beyond the range of single-epoch optical detection.  

\subsection{Synergy between CMB and optical surveys}

As pointed out above, if Planet X can be robustly detected in either a CMB or optical survey alone, it will likely be followed up with targeted observations.  However, there are regimes where the detection and or characterization of Planet X or other outer solar system objects  is greatly helped by a combination of wide-field CMB and optical data.  We summarize the process of identifying Planet X candidates in both datasets in Figures \ref{fig:flow-cmb} and \ref{fig:flow-optical}.

Of particular interest is the case where the parallax of Planet X is too small to be measured by the CMB survey.  The precise distance range at which the parallax becomes unresolvable depends on (a) the resolution of the experiment, (b) the baseline of observation time, (c) the signal-to-noise of the detection.  Higher resolution, longer baseline, and higher signal-to-noise detections would all increase the distance at which a parallax could be resolved.  Very roughly, the threshold of detectable parallax occurs at $d \gtrsim 7000\,{\rm AU}$ for a one arcminute beam or $d \gtrsim 4700\,{\rm AU}$ for the beam of the planned SO telescope and others with a $\sim 6$m primary dish.  These number can vary significantly (factors of two) depending on the search strategy.

In the case that the parallax is not resolvable by the CMB experiment, a planet would appear as an elongated point source, or just as a point source.  Given a catalog of CMB-detected point sources, one could first use multiband information to reject candidates whose spectra are not consistent with thermal emission.  

The remaining catalog of sources would be dominated by dusty galaxies, but could include planetary sources with unresolvable parallax.  These sources could be used to perform a {\it digital tracking} search: assume the individual exposures lie along all potential apparent motion vectors for Planet X (essentially the trajectory due to parallax for distant candidates), and perform an aligned stack for each such trajectory. For a given position on the sky, varying the distance to the source generates a family of possible trajectories.  This allows one to improve the detection limit corresponding to the total survey integration time, i.e. the ``coadd depth'' of the survey.  The technique has long been applied to TNO searches in smaller optical datasets \citep{Tyson:1992,Allen:2001}.   A digital tracking search of the optical survey data would, according to Figure~\ref{fig:signal-noise}, have a good chance of recovering a true Planet X, at the same time determining its orbital properties.  This is the ``sweet spot'' for combined use of wide-field CMB and optical survey data.

The phase space to be searched would be strongly constrained by the CMB data, including its non-detection of parallax. 
If, however, the density of CMB-detected point sources is very high, then the computational burden of digital tracking the CMB sources will become high, eventually approaching the cost of a blind digital tracking search of the full optical survey.  For current sensitivities, the density of extragalactic point sources is roughly $10^{-4}\,{\rm arcmin}^{-2}$.  Assuming point sources can be localized to below a tenth of a beam size, this density corresponds to a filling factor on the sky of about $10^{-6}$.  Therefore, for current CMB surveys, source lists cut the digital tracking load substantially.  However, for more sensitive CMB missions, the density of extragalactic sources is expected to increase dramatically.  For a point source detection threshold of $0.3\,{\rm mJy}$ (i.e. CMB-S4), a reasonable point source density at 220 GHz is $1\,{\rm arcmin}^{-2}$ \citep{Lima:2010}, corresponding to a filling factor of $f = 0.01$.  Note, however, that the number of extragalactic sources can be reduced dramatically by going to lower frequencies.   Since the extragalactic sources are dominated by dust emission, the wavelength dependence of the dust emissivity means that the brightness of these sources changes faster than the $\nu^2$ expected for a planet in the Rayleigh Jeans regime.  Consequently, lower frequency observations can help reduce potential problems with confusion of sources.  Using the 150 GHz band rather than 220 GHz, for instance, is expected to result in a factor of $\sim 100$ reduction in the source density \citep{Lima:2010}, while reducing the signal by a factor of $\sim 2$. 

Another potential regime of interest is the case where the number of sources with parallax measured in the CMB survey is too large to be followed up individually in the optical survey.  While this is unlikely to be the case for planet-sized candidates, CMB surveys may detect many new asteroids and Kuiper belt objects.  If these objects have low albedo, then they may be undetectable in the optical survey.  In this case, a {\it digital tracking} search with the wide-field optical survey would again be fruitful.

\subsection{Thermal properties of detected objects}

Beyond detection, the combination of optical and CMB surveys also offers the prospect of constraining the properties of outer solar system objects.  If the temperature of an object is sufficiently low that the CMB survey observations are not restricted to the Rayleigh-Jeans regime, multi-band observations with the CMB survey can be used to constrain its temperature. Once an object is located,  ALMA may be the best source of flux measurements in the mm range. 

Assuming the parallax motion of the object is also measured (whether with an optical or a CMB survey),  the distance to the object can be constrained.  Given the distance and temperature measurements, the radius of the object as a function of the emissivity can be inferred from the CMB flux measurements via $R(\epsilon) = (4r^2 F_{\rm CMB}/(\sigma \epsilon \pi T^4))^{1/2}$ where $\epsilon$ is the emissivity.  The radius and optical flux can then be used to infer the albedo via $a(\epsilon) = 4\pi r^4 F_{\rm opt}/(L_{\lambda}R^2(\epsilon))$, where $L_{\lambda}$ is the luminosity of the sun in the waveband of interest.  Consequently, the CMB and optical measurements together enable constraints in the albedo-emissivity plane.   Note that measuring a temperature for the detected objects requires that the CMB observations are not entirely in the Rayleigh-Jeans limit; for higher temperatures other instruments will be needed to determine the temperature.

\subsection{Asteroids}

A search for moving objects will image a large number of asteroids, some at extremely high signal-to-noise. The same calculation that was done for planet fluxes can be applied for asteroids, where the internal heat is presumably negligible and the second term in Equation 1 dominates. An asteroid at a distance of 1 AU from Earth and a diameter of 4 km will have an approximate flux at 150 GHz of 0.1 mJy. This same object, assuming an albedo of 0.15, would have an absolute magnitude $H\sim 15$,  hundreds of times brighter than objects detectable with modern surveys such as the Catalina Sky Survey or Pan-Starrs. Therefore, the typical measurable asteroids for a CMB experiment will be those with diameters of many km, of which there are several thousand known objects. This is not the regime of discovering potentially hazardous new objects with typical albedos. CMB surveys will be useful for detecting particularly non-reflective objects and for long-term measurements of rotation curves for tumbling objects and comparing with optical measurements to separate the effects of geometry and reflectivity to infer surface properties. By measuring in the mm, the emissivity should not be strongly affected by details of surface chemistry, also providing a valuable point of comparison with measurements from the NEOWISE program \citep{mainzer2014}.

\section{Outlook}

\begin{figure*}[bh]
\centering
\includegraphics[scale=0.5]{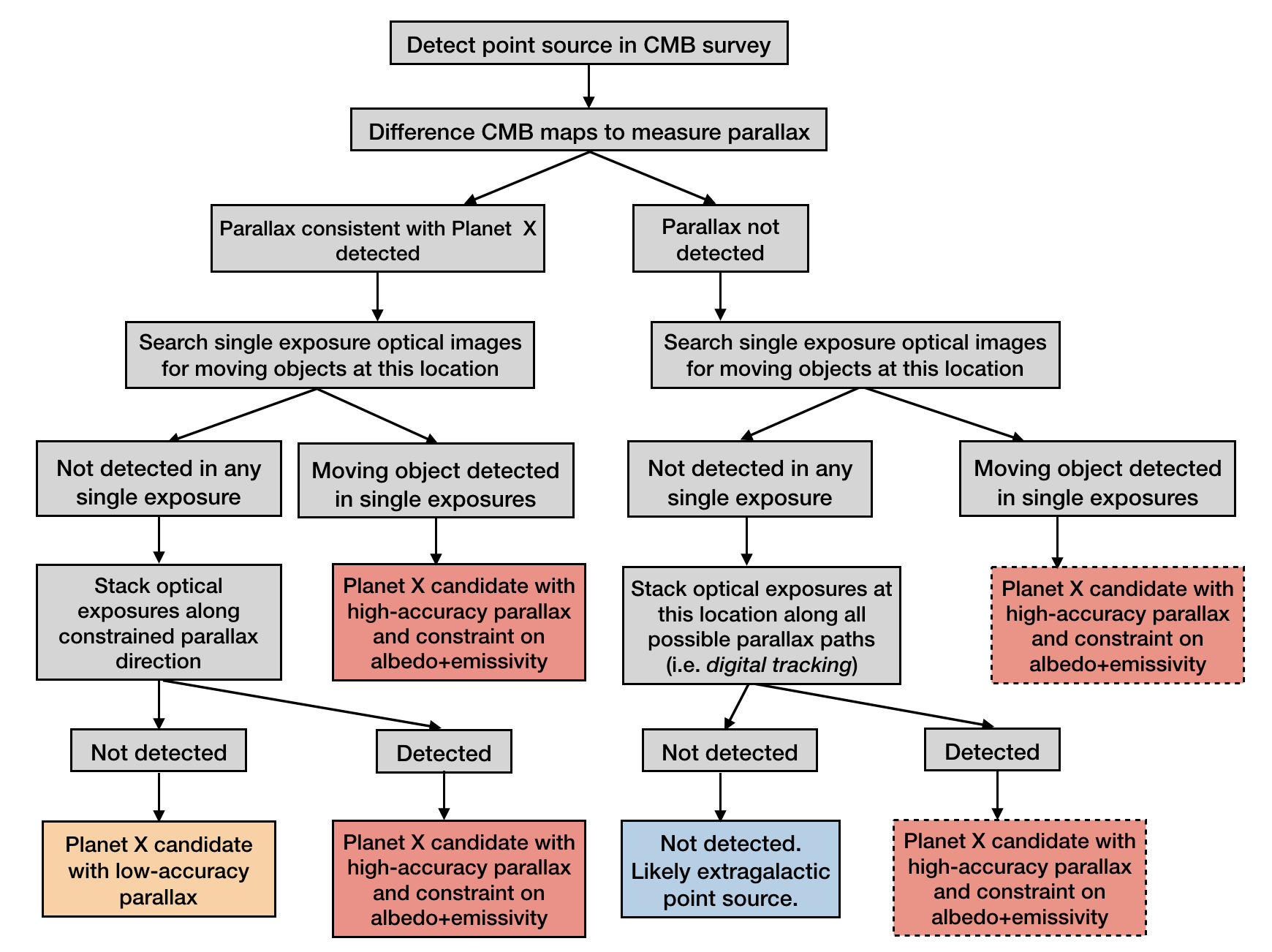}
\caption{
Identification of Planet X candidates starting from the detection of a point source in a CMB survey.  Red boxes indicate the end result of detection of the source and its parallax in both CMB and optical surveys, while the orange box indicates detection in only the CMB survey.  Dashed boxes indicate potential issues of source confusion with extragalactic CMB point sources.  Note that the branch for CMB detected parallax could also be pursued by targeted observations with 8-m class optical telescopes since the number of such detections are likely to be small. 
}
\label{fig:flow-cmb}
\end{figure*}

\begin{figure*}[bh]
\centering
\includegraphics[scale=0.5]{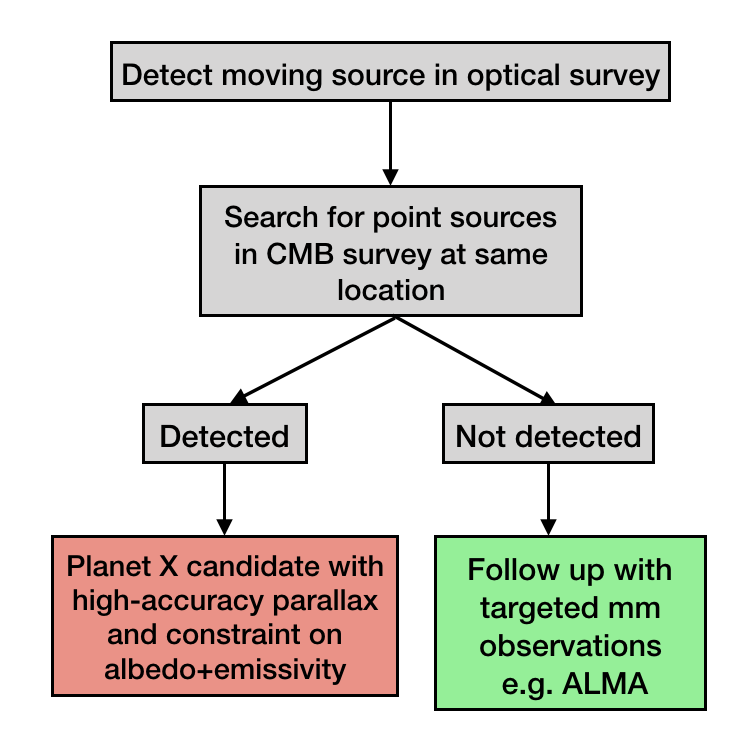}
\caption{
Same as Figure~\ref{fig:flow-cmb}, except now starting from detection in optical survey.
}
\label{fig:flow-optical}
\end{figure*}

We have considered how a combination of data from CMB and optical surveys can be used to improve the prospects for detecting and characterizing a Planet X and other possible outer solar system objects.  As shown in Figure 4, both types of surveys have the signal-to-noise for the detection of Neptune-like and Earth-like planets in interesting regimes. 
 There are several scenarios where the combination of wide field data from {\it both} survey types proves useful:
\begin{enumerate}
    \item The distance to Planet X is sufficiently large, or the baseline of CMB observation times sufficiently short, that its parallax motion cannot be detected in the CMB survey.  The CMB survey then provides a catalog of thermally emitting point sources that can be followed up by the optical survey. For sources too faint to be detected in individual exposures, we discuss how {\it digital tracking} along possible parallax trajectories effectively extends the magnitude limit of the optical survey for Planet X searches.
    \item The optical survey detects many objects with parallax, and the CMB survey is used to measure their thermal emission to characterize their properties such as their albedos.
    \item For objects such as asteroids, the number of objects with parallax detected in the CMB survey could be so large that targeted follow-up with other instruments is prohibitive.   
    \item If a small number of high-likelihood Planet X candidates is identified in data from either survey type,  follow-up could be performed with dedicated instruments. In addition to supporting the follow-up, surveys could provide data on the past positions of the candidates and enable a much faster determination of the orbit.   
\end{enumerate}
Flow charts summarizing the identification process of Planet X candidates in CMB and optical data are shown in Figures \ref{fig:flow-cmb} and \ref{fig:flow-optical}.

A large number of funded and proposed CMB, optical and infrared surveys  are planned for the next decade. At this stage, developing detailed plans for Planet X studies is not feasible for projects that are exploring their survey strategy. This study should help provide some guidance in what would aid planetary studies, though the central goals of the surveys are cosmological. In particular, for CMB surveys it is helpful to have a longer (up to a year) baseline for repeated observations of the entire survey footprint. Follow up studies of detection strategies are merited once the cadences of relevant surveys are know. Finally, we note that the orbital and thermal properties of Planet X candidates are poorly known, in particular its internal heat, albedo and cloud properties. Hence the signal-to-noise estimates discussed above, and even the choice of the best instruments, is subject to revision as we learn more about planetary-sized bodies in the outer solar system.

\vspace{1cm}

{\it Acknowledgements}

We are grateful to Tom Crawford, Cyrille Doux, Mike Jarvis, Matt Lehner,  Mathew Madhavacheril, Renu Malhotra, Surhud More, Sigurd Naess, Megan Schwamb, Lucas Secco and David Spergel for helpful discussions. 

\clearpage

\bibliography{refs.bib}

\begin{thebibliography}{}
\expandafter\ifx\csname natexlab\endcsname\relax\def\natexlab#1{#1}\fi
\providecommand{\url}[1]{\href{#1}{#1}}

\bibitem[{{Abazajian} {et~al.}(2016){Abazajian}, {Adshead}, {Ahmed}, {Allen},
  {Alonso}, {Arnold}, {Baccigalupi}, {Bartlett}, {Battaglia}, {Benson},
  {Bischoff}, {Borrill}, {Buza}, {Calabrese}, {Caldwell}, {Carlstrom}, {Chang},
  {Crawford}, {Cyr-Racine}, {De Bernardis}, {de Haan}, {di Serego Alighieri},
  {Dunkley}, {Dvorkin}, {Errard}, {Fabbian}, {Feeney}, {Ferraro}, {Filippini},
  {Flauger}, {Fuller}, {Gluscevic}, {Green}, {Grin}, {Grohs}, {Henning},
  {Hill}, {Hlozek}, {Holder}, {Holzapfel}, {Hu}, {Huffenberger}, {Keskitalo},
  {Knox}, {Kosowsky}, {Kovac}, {Kovetz}, {Kuo}, {Kusaka}, {Le Jeune}, {Lee},
  {Lilley}, {Loverde}, {Madhavacheril}, {Mantz}, {Marsh}, {McMahon},
  {Meerburg}, {Meyers}, {Miller}, {Munoz}, {Nguyen}, {Niemack}, {Peloso},
  {Peloton}, {Pogosian}, {Pryke}, {Raveri}, {Reichardt}, {Rocha}, {Rotti},
  {Schaan}, {Schmittfull}, {Scott}, {Sehgal}, {Shandera}, {Sherwin}, {Smith},
  {Sorbo}, {Starkman}, {Story}, {van Engelen}, {Vieira}, {Watson}, {Whitehorn},
  \& {Kimmy Wu}}]{Abazajian:2016}
{Abazajian}, K.~N., {Adshead}, P., {Ahmed}, Z., {et~al.} 2016, arXiv e-prints,
  arXiv:1610.02743

\bibitem[{{Allen} {et~al.}(2001){Allen}, {Bernstein}, \&
  {Malhotra}}]{Allen:2001}
{Allen}, R.~L., {Bernstein}, G.~M., \& {Malhotra}, R. 2001, \apj, 549, L241

\bibitem[{{Batygin} \& {Brown}(2016)}]{Batygin:2016}
{Batygin}, K., \& {Brown}, M.~E. 2016, \aj, 151, 22

\bibitem[{{Benson} {et~al.}(2014){Benson}, {Ade}, {Ahmed}, {Allen}, {Arnold},
  {Austermann}, {Bender}, {Bleem}, {Carlstrom}, {Chang}, {Cho}, {Cliche},
  {Crawford}, {Cukierman}, {de Haan}, {Dobbs}, {Dutcher}, {Everett}, {Gilbert},
  {Halverson}, {Hanson}, {Harrington}, {Hattori}, {Henning}, {Hilton},
  {Holder}, {Holzapfel}, {Irwin}, {Keisler}, {Knox}, {Kubik}, {Kuo}, {Lee},
  {Leitch}, {Li}, {McDonald}, {Meyer}, {Montgomery}, {Myers}, {Natoli},
  {Nguyen}, {Novosad}, {Padin}, {Pan}, {Pearson}, {Reichardt}, {Ruhl},
  {Saliwanchik}, {Simard}, {Smecher}, {Sayre}, {Shirokoff}, {Stark}, {Story},
  {Suzuki}, {Thompson}, {Tucker}, {Vanderlinde}, {Vieira}, {Vikhlinin}, {Wang},
  {Yefremenko}, \& {Yoon}}]{Benson:2014}
{Benson}, B.~A., {Ade}, P.~A.~R., {Ahmed}, Z., {et~al.} 2014, in Society of
  Photo-Optical Instrumentation Engineers (SPIE) Conference Series, Vol. 9153,
  Millimeter, Submillimeter, and Far-Infrared Detectors and Instrumentation for
  Astronomy VII, 91531P

\bibitem[{{Bernardinelli} {et~al.}(2019){Bernardinelli}, {Bernstein}, {Sako},
  \& {the Dark Energy Survey Collaboration}}]{Pedro:2019}
{Bernardinelli}, P., {Bernstein}, G., {Sako}, M., \& {the Dark Energy Survey
  Collaboration}. 2019, in preparation

\bibitem[{Brown \& Batygin(2016)}]{Brown:2016}
Brown, M.~E., \& Batygin, K. 2016, The Astrophysical Journal Letters, 824, L23.
\newblock \url{http://stacks.iop.org/2041-8205/824/i=2/a=L23}

\bibitem[{Carlstrom {et~al.}(2011)Carlstrom, Ade, Aird, Benson, Bleem, Busetti,
  Chang, Chauvin, Cho, Crawford, Crites, Dobbs, Halverson, Heimsath, Holzapfel,
  Hrubes, Joy, Keisler, Lanting, Lee, Leitch, Leong, Lu, Lueker, Luong-Van,
  McMahon, Mehl, Meyer, Mohr, Montroy, Padin, Plagge, Pryke, Ruhl, Schaffer,
  Schwan, Shirokoff, Spieler, Staniszewski, Stark, Tucker, Vanderlinde, Vieira,
  \& Williamson}]{Carlstrom:2011}
Carlstrom, J.~E., Ade, P. A.~R., Aird, K.~A., {et~al.} 2011, Publications of
  the Astronomical Society of the Pacific, 123, 568.
\newblock \url{http://stacks.iop.org/1538-3873/123/i=903/a=568}

\bibitem[{{Cowan} {et~al.}(2016){Cowan}, {Holder}, \& {Kaib}}]{Cowan:2016}
{Cowan}, N.~B., {Holder}, G., \& {Kaib}, N.~A. 2016, \apjl, 822, L2

\bibitem[{{Flaugher}(2005)}]{Flaugher:2005}
{Flaugher}, B. 2005, International Journal of Modern Physics A, 20, 3121

\bibitem[{{Galitzki} {et~al.}(2018){Galitzki}, {Ali}, {Arnold}, {Ashton},
  {Austermann}, {Baccigalupi}, {Baildon}, {Barron}, {Beall}, {Beckman},
  {Bruno}, {Bryan}, {Calisse}, {Chesmore}, {Chinone}, {Choi}, {Coppi},
  {Crowley}, {Crowley}, {Cukierman}, {Devlin}, {Dicker}, {Dober}, {Duff},
  {Dunkley}, {Fabbian}, {Gallardo}, {Gerbino}, {Goeckner- Wald}, {Golec},
  {Gudmundsson}, {Healy}, {Henderson}, {Hill}, {Hilton}, {Ho}, {Howe},
  {Hubmayr}, {Jeong}, {Keating}, {Koopman}, {Kiuchi}, {Kusaka}, {Lashner},
  {Lee}, {Li}, {Limon}, {Lungu}, {Matsuda}, {Mauskopf}, {May}, {McCallum},
  {McMahon}, {Nati}, {Niemack}, {Orlowski-Scherer}, {Parshley}, {Piccirillo},
  {Sathyanarayana Rao}, {Raum}, {Salatino}, {Seibert}, {Sierra},
  {Silva-Feaver}, {Simon}, {Staggs}, {Stevens}, {Suzuki}, {Teply}, {Thornton},
  {Tsai}, {Ullom}, {Vavagiakis}, {Vissers}, {Westbrook}, {Wollack}, {Xu}, \&
  {Zhu}}]{Galitzki:2018}
{Galitzki}, N., {Ali}, A., {Arnold}, K.~S., {et~al.} 2018, in Society of
  Photo-Optical Instrumentation Engineers (SPIE) Conference Series, Vol. 10708,
  Millimeter, Submillimeter, and Far-Infrared Detectors and Instrumentation for
  Astronomy IX, 1070804

\bibitem[{{Henderson} {et~al.}(2016){Henderson}, {Allison}, {Austermann},
  {Baildon}, {Battaglia}, {Beall}, {Becker}, {De Bernardis}, {Bond},
  {Calabrese}, {Choi}, {Coughlin}, {Crowley}, {Datta}, {Devlin}, {Duff},
  {Dunkley}, {D{\"u}nner}, {van Engelen}, {Gallardo}, {Grace}, {Hasselfield},
  {Hills}, {Hilton}, {Hincks}, {Hlo{\^z}ek}, {Ho}, {Hubmayr}, {Huffenberger},
  {Hughes}, {Irwin}, {Koopman}, {Kosowsky}, {Li}, {McMahon}, {Munson}, {Nati},
  {Newburgh}, {Niemack}, {Niraula}, {Page}, {Pappas}, {Salatino}, {Schillaci},
  {Schmitt}, {Sehgal}, {Sherwin}, {Sievers}, {Simon}, {Spergel}, {Staggs},
  {Stevens}, {Thornton}, {Van Lanen}, {Vavagiakis}, {Ward}, \&
  {Wollack}}]{Henderson:2016}
{Henderson}, S.~W., {Allison}, R., {Austermann}, J., {et~al.} 2016, Journal of
  Low Temperature Physics, 184, 772

\bibitem[{{Lima} {et~al.}(2010){Lima}, {Jain}, {Devlin}, \&
  {Aguirre}}]{Lima:2010}
{Lima}, M., {Jain}, B., {Devlin}, M., \& {Aguirre}, J. 2010, \apj, 717, L31

\bibitem[{{LSST Dark Energy Science Collaboration}(2012)}]{LSST:2012}
{LSST Dark Energy Science Collaboration}. 2012, arXiv e-prints, arXiv:1211.0310

\bibitem[{{Mainzer} {et~al.}(2014){Mainzer}, {Bauer}, {Cutri}, {Grav},
  {Masiero}, {Beck}, {Clarkson}, {Conrow}, {Dailey}, {Eisenhardt}, {Fabinsky},
  {Fajardo-Acosta}, {Fowler}, {Gelino}, {Grillmair}, {Heinrichsen}, {Kendall},
  {Kirkpatrick}, {Liu}, {Masci}, {McCallon}, {Nugent}, {Papin}, {Rice},
  {Royer}, {Ryan}, {Sevilla}, {Sonnett}, {Stevenson}, {Thompson}, {Wheelock},
  {Wiemer}, {Wittman}, {Wright}, \& {Yan}}]{mainzer2014}
{Mainzer}, A., {Bauer}, J., {Cutri}, R.~M., {et~al.} 2014, \apj, 792, 30

\bibitem[{{Malhotra} {et~al.}(2016){Malhotra}, {Volk}, \&
  {Wang}}]{Malhotra:2016}
{Malhotra}, R., {Volk}, K., \& {Wang}, X. 2016, \apjl, 824, L22

\bibitem[{{Rowe} {et~al.}(2008){Rowe}, {Matthews}, {Seager}, {Miller-Ricci},
  {Sasselov}, {Kuschnig}, {Guenther}, {Moffat}, {Rucinski}, {Walker}, \&
  {Weiss}}]{Rowe:2008}
{Rowe}, J.~F., {Matthews}, J.~M., {Seager}, S., {et~al.} 2008, \apj, 689, 1345

\bibitem[{{Sheppard} {et~al.}(2018){Sheppard}, {Trujillo}, {Tholen}, \&
  {Kaib}}]{Sheppard:2018}
{Sheppard}, S., {Trujillo}, C., {Tholen}, D., \& {Kaib}, N. 2018, ArXiv
  e-prints, arXiv:1810.00013

\bibitem[{{Sievers} {et~al.}(2013){Sievers}, {Hlozek}, {Nolta}, {Acquaviva},
  {Addison}, {Ade}, {Aguirre}, {Amiri}, {Appel}, {Barrientos}, {Battistelli},
  {Battaglia}, {Bond}, {Brown}, {Burger}, {Calabrese}, {Chervenak}, {Crichton},
  {Das}, {Devlin}, {Dicker}, {Bertrand Doriese}, {Dunkley}, {D{\"u}nner},
  {Essinger-Hileman}, {Faber}, {Fisher}, {Fowler}, {Gallardo}, {Gordon},
  {Gralla}, {Hajian}, {Halpern}, {Hasselfield}, {Hern{\'a}ndez-Monteagudo},
  {Hill}, {Hilton}, {Hilton}, {Hincks}, {Holtz}, {Huffenberger}, {Hughes},
  {Hughes}, {Infante}, {Irwin}, {Jacobson}, {Johnstone}, {Baptiste Juin},
  {Kaul}, {Klein}, {Kosowsky}, {Lau}, {Limon}, {Lin}, {Louis}, {Lupton},
  {Marriage}, {Marsden}, {Martocci}, {Mauskopf}, {McLaren}, {Menanteau},
  {Moodley}, {Moseley}, {Netterfield}, {Niemack}, {Page}, {Page}, {Parker},
  {Partridge}, {Plimpton}, {Quintana}, {Reese}, {Reid}, {Rojas}, {Sehgal},
  {Sherwin}, {Schmitt}, {Spergel}, {Staggs}, {Stryzak}, {Swetz}, {Switzer},
  {Thornton}, {Trac}, {Tucker}, {Uehara}, {Visnjic}, {Warne}, {Wilson},
  {Wollack}, {Zhao}, \& {Zunckel}}]{Sievers:2013}
{Sievers}, J.~L., {Hlozek}, R.~A., {Nolta}, M.~R., {et~al.} 2013, Journal of
  Cosmology and Astro-Particle Physics, 2013, 060

\bibitem[{{Stacey} {et~al.}(2018){Stacey}, {Aravena}, {Basu}, {Battaglia},
  {Beringue}, {Bertoldi}, {Bond}, {Breysse}, {Bustos}, {Chapman}, {Chung},
  {Cothard}, {Erler}, {Fich}, {Foreman}, {Gallardo}, {Giovanelli}, {Graf},
  {Haynes}, {Herrera-Camus}, {Herter}, {Hlo{\v{z}}ek}, {Johnstone}, {Keating},
  {Magnelli}, {Meerburg}, {Meyers}, {Murray}, {Niemack}, {Nikola}, {Nolta},
  {Parshley}, {Riechers}, {Schilke}, {Scott}, {Stein}, {Stevens}, {Stutzki},
  {Vavagiakis}, \& {Viero}}]{Stacey:2018}
{Stacey}, G.~J., {Aravena}, M., {Basu}, K., {et~al.} 2018, arXiv e-prints,
  arXiv:1807.04354

\bibitem[{{Stevens} {et~al.}(2018){Stevens}, {Goeckner-Wald}, {Keskitalo},
  {McCallum}, {Ali}, {Borrill}, {Brown}, {Chinone}, {Gallardo}, {Kusaka},
  {Lee}, {McMahon}, {Niemack}, {Page}, {Puglisi}, {Salatino}, {Mak}, {Teply},
  {Thomas}, {Vavagiakis}, {Wollack}, {Xu}, \& {Zhu}}]{Stevens:2018}
{Stevens}, J.~R., {Goeckner-Wald}, N., {Keskitalo}, R., {et~al.} 2018, in
  Society of Photo-Optical Instrumentation Engineers (SPIE) Conference Series,
  Vol. 10708, Millimeter, Submillimeter, and Far-Infrared Detectors and
  Instrumentation for Astronomy IX, 1070841

\bibitem[{{Story} {et~al.}(2013){Story}, {Reichardt}, {Hou}, {Keisler}, {Aird},
  {Benson}, {Bleem}, {Carlstrom}, {Chang}, {Cho}, {Crawford}, {Crites}, {de
  Haan}, {Dobbs}, {Dudley}, {Follin}, {George}, {Halverson}, {Holder},
  {Holzapfel}, {Hoover}, {Hrubes}, {Joy}, {Knox}, {Lee}, {Leitch}, {Lueker},
  {Luong-Van}, {McMahon}, {Mehl}, {Meyer}, {Millea}, {Mohr}, {Montroy},
  {Padin}, {Plagge}, {Pryke}, {Ruhl}, {Sayre}, {Schaffer}, {Shaw}, {Shirokoff},
  {Spieler}, {Staniszewski}, {Stark}, {van Engelen}, {Vanderlinde}, {Vieira},
  {Williamson}, \& {Zahn}}]{Story:2013}
{Story}, K.~T., {Reichardt}, C.~L., {Hou}, Z., {et~al.} 2013, \apj, 779, 86

\bibitem[{{Swetz} {et~al.}(2011){Swetz}, {Ade}, {Amiri}, {Appel},
  {Battistelli}, {Burger}, {Chervenak}, {Devlin}, {Dicker}, {Doriese},
  {D{\"u}nner}, {Essinger-Hileman}, {Fisher}, {Fowler}, {Halpern},
  {Hasselfield}, {Hilton}, {Hincks}, {Irwin}, {Jarosik}, {Kaul}, {Klein},
  {Lau}, {Limon}, {Marriage}, {Marsden}, {Martocci}, {Mauskopf}, {Moseley},
  {Netterfield}, {Niemack}, {Nolta}, {Page}, {Parker}, {Staggs}, {Stryzak},
  {Switzer}, {Thornton}, {Tucker}, {Wollack}, \& {Zhao}}]{Swetz2011}
{Swetz}, D.~S., {Ade}, P.~A.~R., {Amiri}, M., {et~al.} 2011, \apjs, 194, 41

\bibitem[{{Trilling} {et~al.}(2018){Trilling}, {Bellm}, \&
  {Malhotra}}]{Trilling:2018}
{Trilling}, D.~E., {Bellm}, E.~C., \& {Malhotra}, R. 2018, \aj, 155, 243

\bibitem[{{Trujillo} \& {Sheppard}(2014)}]{Trujillo:2014}
{Trujillo}, C.~A., \& {Sheppard}, S.~S. 2014, \nat, 507, 471

\bibitem[{{Tyson} {et~al.}(1992){Tyson}, {Guhathakurta}, {Bernstein}, \&
  {Hut}}]{Tyson:1992}
{Tyson}, J.~A., {Guhathakurta}, P., {Bernstein}, G.~M., \& {Hut}, P. 1992, in
  Bulletin of the American Astronomical Society, Vol.~24, American Astronomical
  Society Meeting Abstracts, 1127

\bibitem[{{Van Tilburg} {et~al.}(2018){Van Tilburg}, {Taki}, \&
  {Weiner}}]{vanTilburg:2018}
{Van Tilburg}, K., {Taki}, A.-M., \& {Weiner}, N. 2018, Journal of Cosmology
  and Astro-Particle Physics, 2018, 041

\bibitem[{Volk \& Malhotra(2017)}]{Volk:2017}
Volk, K., \& Malhotra, R. 2017, The Astronomical Journal, 154, 62.
\newblock \url{http://stacks.iop.org/1538-3881/154/i=2/a=62}

\bibitem[{{Weiss} \& {Marcy}(2014)}]{Weiss:2013}
{Weiss}, L.~M., \& {Marcy}, G.~W. 2014, \apjl, 783, L6

\bibitem[{{Weryk} {et~al.}(2016){Weryk}, {Lilly}, {Chastel}, {Denneau},
  {Jedicke}, {Magnier}, {Wainscoat}, {Chambers}, {Flewelling}, {Huber},
  {Waters}, \& {PS1 Builders}}]{Weryk:2016}
{Weryk}, R.~J., {Lilly}, E., {Chastel}, S., {et~al.} 2016, arXiv e-prints,
  arXiv:1607.04895

\end{thebibliography}

%
\end{document}